\documentclass[aps,pra,twocolumn,superscriptaddress,groupedaddress,floatfix]{revtex4}  % for review and submission

\usepackage{graphicx}  % needed for figures
\usepackage{bm}

\usepackage{graphicx}
\usepackage{dcolumn}
\usepackage{bm}
\usepackage{latexsym}

\usepackage{amssymb,amsmath}
\usepackage{amsthm}
\usepackage{amsfonts}
\usepackage{avant}
\usepackage{textcomp}

\usepackage{amsmath}
\usepackage{dcolumn}
\usepackage{epsfig}
\usepackage{graphicx}
\usepackage{latexsym}

\usepackage{epstopdf}
\usepackage{rotating}

\usepackage{supertabular}
\usepackage{booktabs}
\usepackage{multirow}

\usepackage{siunitx}
\usepackage{xcolor}

\usepackage{float}

\begin{document}

%\preprint{}

\title{Isotope shift spectroscopy of the ${^1}S_0 \rightarrow {^3}P_1$ and ${^1}S_0 \rightarrow {^3}P_0$ transitions in strontium}


%\author{}
%\thanks{}
%\affiliation{}

\author{Hirokazu Miyake}
\author{Neal C. Pisenti}
\author{Peter K. Elgee}
\author{Ananya Sitaram}
\author{Gretchen K. Campbell}
\affiliation{Joint Quantum Institute, University of Maryland and National Institute of Standards and Technology, College Park, Maryland 20742, USA}


%\date{\today}

\begin{abstract}
Isotope shift spectroscopy with narrow optical transitions provides a benchmark for atomic structure calculations and has also been proposed as a way to constrain theories predicting physics beyond the Standard Model. 
Here, we have measured frequency shifts of the ${^1}S_0 \rightarrow {^3}P_1$ and ${^1}S_0 \rightarrow {^3}P_0$ transitions between all stable isotopes of strontium relative to ${^{88}}$Sr.
This includes the first reported measurements of the ${^1}S_0 \rightarrow {^3}P_0$ isotope shift of ${^{88}}$Sr-${^{86}}$Sr and ${^{88}}$Sr-${^{84}}$Sr.  
Using the isotope shift measurements of the two transitions a King plot analysis is performed, revealing a non-linearity in the measured values.
\end{abstract}

%\pacs{}
%\keywords{}

\maketitle

\section{Introduction}
Isotope shifts of atomic transition frequencies arise due to the difference in neutron numbers for different isotopes with the same atomic number.
For a given element, these shifts can be systematically analyzed using a King plot, which elucidates the contributions of the field and mass shifts~\cite{king84}.
The King plot is typically expected to be linear, and the experimentally determined value of the slope provides a good benchmark for theoretical predictions~\cite{flambaum18}. Any deviations from linearity as was observed in Sm~\cite{griffith81} and Ba~\cite{dammalapati09}, or between predicted and experimentally measured values of the slope as was observed in Ca$^+$~\cite{shi17}, are important for refining atomic structure calculations~\cite{naze15}.
Furthermore, recent theoretical proposals have suggested that linearity in King plots could be used to put constraints on higher-order effects on isotope shifts or on physics beyond the Standard Model~\cite{frugiuele17,berengut18}. 
Strontium has many favorable properties for studying isotope shifts, including an abundance of stable isotopes and very narrow optical transitions~\cite{stellmer14}. In addition, prior theoretical work has proposed the measurement of strontium isotope shifts as a promising probe of new physics~\cite{frugiuele17,berengut18}.

Strontium has four stable isotopes: three bosons ($^{88}$Sr, $^{86}$Sr, and $^{84}$Sr), and one fermion ($^{87}$Sr).
Mixing between the singlet and triplet fine structure manifolds leads to narrow-linewidth optical transitions, and these transitions have found use in both strontium and other alkaline-earth-(like) atom experiments~\cite{daley11,he19}.  In particular for strontium, the ${^1}S_0 \rightarrow {^3}P_1$ intercombination-line transition at 689~nm (linewidth $\Gamma/2\pi = 7.4$~kHz) is used during laser cooling to operate a narrow-line magneto-optical trap (MOT)~\cite{katori99,stellmer14}, and the even narrower ${^1}S_0 \rightarrow {^3}P_0$ clock transition at 698~nm ($\Gamma/2\pi \sim$~mHz) is the foundation for state-of-the-art optical clocks operating at a precision at the $10^{-18}$ level~\cite{bloom14,nicholson15,campbell17}. 
The clock transition is strictly forbidden by angular momentum considerations, but becomes weakly allowed via hyperfine mixing in $^{87}$Sr or by application of an external field for the bosonic isotopes~\cite{taichenachev06}.

While the ${^1}S_0 \rightarrow {^3}P_0$ clock transition has been extensively studied in $^{87}$Sr and $^{88}$Sr~\cite{bloom14,nicholson15,campbell17,taichenachev06,baillard07,akatsuka08,morzynski15,takano17}, there have been no previous measurements of the transition in either $^{86}$Sr or $^{84}$Sr as far as we know~\cite{ludlow15}.
Here we report the first isotope shift spectroscopy measurements of the clock transition for both $^{84}$Sr and $^{86}$Sr relative to the most abundant isotope $^{88}$Sr. 
Furthermore, we measure all isotope shifts of the ${^1}S_0 \rightarrow {^3}P_1$ intercombination-line transition relative to $^{88}$Sr, permitting the first King plot analysis of strontium for these two transitions.
Given the very narrow linewidths involved, extensions of this work could place stringent experimental constraints on the King linearity, ruling out candidate theories for physics beyond the Standard Model or benchmarking state-of-the-art atomic structure calculations.

\section{Experimental procedure}
All of the isotope shift spectroscopy was performed using laser-cooled strontium atoms at temperatures of a few $\mu$K, held in an optical dipole trap (ODT).  After applying the spectroscopy light, we monitored atom loss by performing absorption imaging. 

\begin{table*}[htbp]
\begin{center}
\begin{tabular}{c c c c c c c}
  \hline \hline	
  & 88 & \multicolumn{3}{c}{87} & 86 & 84 \\
  \cline{3-5}
  Systematic Shift (kHz) &  & $F'=7/2$ & $F'=9/2$ & $F'=11/2$ &  &    \\ \hline
  Density & $1.7 \pm 1.7$ & $-34.1 \pm 14.5$ & $-51.9 \pm 26.8$ & $-43.3 \pm 15.6$ & $5.1 \pm 3.4$ & $-1.4 \pm 4.3$ \\
  Recoil & $4.8 \pm (<0.1)$ & $4.8 \pm (<0.1)$ & $4.8 \pm (<0.1)$ & $4.8 \pm (<0.1)$ & $4.9 \pm (<0.1)$ & $5.0 \pm (<0.1)$ \\
  \hline
  Total & $6.5 \pm 1.7$ & $-29.3 \pm 14.5$ & $-47.1 \pm 26.8$ & $-38.5 \pm 15.6$ & $10.0 \pm 3.4$ & $3.6 \pm 4.3$ \\
  \hline \hline
\end{tabular}
\caption{Measured systematic frequency shifts and uncertainties for the ${^1}S_0 \rightarrow {^3}P_1$ transition.  
Uncertainties indicate one standard deviation.
\label{table:intercombsystematics}}
\end{center}
\end{table*}
The laser lights used for spectroscopy of both the ${^1}S_0 \rightarrow {^3}P_1$ and ${^1}S_0 \rightarrow {^3}P_0$ lines were generated using two home-built external-cavity diode lasers based on the design in Ref.~\cite{cook12}. 
The frequency of the 689-nm laser was stabilized via an optical phase-locked loop~\cite{appel09} to the master laser of the 689-nm narrow-line MOT system.
The master 689-nm laser was locked using the Pound-Drever-Hall (PDH) method~\cite{drever83,black01} to a cavity constructed from ultra-low expansion (ULE) glass and housed in a temperature-stabilized vacuum chamber.
To stabilize the frequency of the 698-nm laser, we passed a few percent of the light through a wide bandwidth electro-optic modulator and locked the first phase-modulated sideband via PDH to a second, independent ULE cavity~\cite{thorpe08}. This cavity was housed in a separate temperature-stabilized, acoustically-isolated vacuum chamber.
These locking schemes for both lasers allowed us the flexibility to shift the frequency of either laser to span the isotope shifts of its respective transition.
For both the 689-nm and 698-nm lasers, the light was referenced to a frequency comb (Menlo System FC1500-250-ULN~\cite{nist_disclaimer}) to account for long-term drift and provide a frequency reference.
Fine frequency control of each laser beam was achieved by adjusting the drive of an acousto-optic modulator, which was also used to stabilize the intensity of the spectroscopy pulse.
The spectroscopy laser linewidth was characterized by locking independent 689-nm lasers to each ULE cavity. A heterodyne beatnote at 689~nm between the two separate lasers was measured to be approximately 200~Hz wide, which bounds the expected spectral performance of both systems.
 
The remainder of the apparatus used for the spectroscopy has been described in detail previously~\cite{barker15}.
Laser cooling of all isotopes proceeds according to well-established techniques~\cite{stellmer14}, with a MOT first operating on the broad ${^1}S_0 \rightarrow {^1}P_1$
transition at 461 nm, followed by a narrow-line MOT operating on the 689-nm intercombination-line transition ${^1}S_0 \rightarrow {^3}P_1$.
For all isotopes, temperatures in the narrow-line MOT are typically a few $\mu$K, low enough to efficiently transfer the atoms into a single beam, far-detuned ODT at 1064~nm.
Typical temperatures in the ODT are \{$2.9$ $\mu$K, $2.2$ $\mu$K, $1.1$ $\mu$K $2.7$ $\mu$K\} and typical atom numbers are \{$1$, $0.1$, $0.5$, $0.2$\}$\times 10^6$ for \{${^{88}}$Sr, ${^{87}}$Sr, ${^{86}}$Sr, ${^{84}}$Sr\} respectively, with trap frequencies \{$\omega_x$, $\omega_y$, $\omega_z$\}$/(2\pi) = \{50, 4, 495\}$~Hz in the horizontal, axial, and vertical directions respectively. For the bosons, the variation in atom number is mostly due to the difference in the natural abundance of each isotope, whereas for the fermionic isotope the atom number is also limited by the additional complexity of the narrow-line MOT~\cite{mukaiyama03}.

\section{Measurement of the ${^1}S_0 \rightarrow {^3}P_1$ isotope shifts}
After loading the atoms into the ODT, the magnetic field was set to 
$0.05$~mT
 (0.5~G) to resolve the ${^3}P_1 (m=0)$ state for the even (bosonic) isotopes.
For the odd (fermionic) isotope, which has hyperfine structure, the magnetic field was set to zero, meaning that the Zeeman splitting was not detectable within the lineshape.
The strength of the magnetic field was calibrated by addressing the ${^1}S_0(m = 0) \rightarrow {^3}P_1 (m' = 1)$ transition of ${^{88}}$Sr, for which the Zeeman shift is known~\cite{katori99}.
To eliminate the effect of AC Stark shifts from the ODT we implemented a stroboscopic procedure where, with a typical period of 500 $\mu$s, the ODT was turned on and off with a duty cycle of 50\% (duration of 250 $\mu$s), and applied the 689-nm probe laser when the ODT was off, similar to the procedure used in Refs.~\cite{borkowski14,reschovsky18}.
The spectroscopy light was used to induce atom loss from the trap through light scattering and subsequent recoil, which is primarily an incoherent process where there is absence of coherence between the pulses of the stroboscopic method with the switching of the ODT.
The 689-nm spectroscopy beam was aligned at an angle of approximately $45^\circ$ with respect to the ODT, both in the horizontal plane. The spectroscopy beam was collimated with a $1/e^2$ beam waist of $1.25$ mm in the horizontal direction and $1.71$ mm in the vertical direction at the position of the atoms.
The polarization of the spectroscopy beam was set to be linear along the direction of the magnetic field.
The total illumination duration used for spectroscopy was set to between 1 ms and 15 ms, corresponding to multiple stroboscopic pulses, and the peak optical intensity was at most 0.1 mW/cm$^2$ ($I_{\rm sat} = 3$ $\mu$W/cm$^2$). These values were chosen to ensure atom loss of approximately 50\%.
After the spectroscopy pulse was completed, the atoms were released from the ODT and we performed absorption imaging on the $^{1}S_0 \rightarrow {^1}P_1$ transition to measure atom loss as a function of the spectroscopy laser frequency.

For all four isotopes, data was taken across several days and referenced to the frequency comb. 
Then the frequencies were averaged to obtain a single line center for each isotope. 
A final isotope shift was found by subtracting the measured absolute frequencies relative to $^{88}$Sr, and the total errors were added in quadrature.
For the $^{87}$Sr isotope shift, we weight the measurements of each excited-state hyperfine manifold $F' \in \{11/2,9/2,7/2\}$ to find the nominally unshifted line center in the absence of the hyperfine interaction~\cite{atomicStructure}.
However, it is important to note that this model fits three parameters (hyperfine $A$ and $B$ coefficients and an unshifted line center) from three isotope shifts, and thus is completely determined by the available data~\cite{hfcoeff}. 
A more accurate theory of higher order shifts from other fine structure levels will be necessary to assign a more accurate isotope shift for $^{87}$Sr. 
This is currently an area of ongoing theoretical research~\cite{beloy08}. 

To calculate the final value for the isotope shift, we also evaluated systematic effects, as summarized in Table~\ref{table:intercombsystematics}. Since many of the systematic effects are common to both isotopes, and the isotope shift is found from a difference in those frequencies, many potential systematic effects are common mode and cancel to a high degree. This is particularly true for the even isotopes, where there is no hyperfine structure.
For example, even though a magnetic field is applied during the spectroscopy pulse for the even isotopes, the Zeeman shift is identical to within our experimental uncertainties, and does not lead to a correction to the final isotope shift.  Therefore, as shown in Table~\ref{table:intercombsystematics}, the remaining systematic effects are those that are not common mode: the density shift and recoil shift.

The density shift arises due to the different scattering lengths and atom numbers between different isotopes in our experiment. The cumulative effect is a non-zero differential density shift to the final isotope shift value. 
We experimentally determined this density shift for each isotope by measuring the line center at different atom numbers while keeping all other parameters the same.
A linear fit allowed us to extrapolate from our operating atom number to a nominal ``zero-density'' frequency, yielding the systematic density shift shown in Table~\ref{table:intercombsystematics}.
The photon recoil shift~\cite{ido03} was also accounted for and was calculated from known physical quantities.

To first order, the ${^{1}}S_0 \rightarrow {^{3}}P_1$ transition is magnetic field insensitive, and our measurements were performed at a low magnetic field of 
$0.05$~mT
(0.5 G) for the bosons and zero magnetic field for the fermion. Therefore both the first and second order Zeeman shifts were negligible at our level of accuracy. The stroboscopic procedure described above removed any AC Stark shifts due to the 1064-nm trapping beam.
Finally, since the intensity in the 689-nm spectrosopy pulse was low (at most 0.1 mW/cm$^2$) and the probe times were short (a few ms), systematic shifts from the probe pulse were below our experimental uncertainty.

\begin{table}[htbp]
\begin{center}
\begin{tabular}{c c c}
  \hline \hline	
  Isotope Shift (kHz) & {${^{1}}S_0 \rightarrow {^{3}}P_1$}  & {${^{1}}S_0 \rightarrow {^{3}}P_0$}   \\ \hline
  88-84 & $351495.8 \pm 0.3 \pm  4.6$ &  $349 656 \pm 1 \pm 10$ \\
  88-86 & $163818.7 \pm 0.3 \pm  3.8$ &  $162 939 \pm 2 \pm 11$ \\
  88-87 & $62186.5  \pm 0.6 \pm 11.7$ &  $ 62 171  \pm 1 \pm 23$ \\
  88-87 ($F'=7/2$) & $-1351933.1  \pm 2.1 \pm 14.6$ &   \\
  88-87 ($F'=9/2$) & $-221676.6  \pm 0.4 \pm 26.9$ &   \\
  88-87 ($F'=11/2$) & $1241485.8  \pm 0.3 \pm 15.7$ &   \\
  \hline \hline
\end{tabular}
\caption{Measured isotope shifts relative to $^{88}$Sr.
For $^{87}$Sr (${^{1}}S_0 \rightarrow {^{3}}P_1$), contributions from the three excited-state hyperfine manifolds are weighted to establish the fine-structure line center.
Uncertainties are one standard deviation and indicate statistical and systematic uncertainties.
\label{table:isotopeshift}}
\end{center}
\end{table}

After applying corrections for the systematic effects, the final values for ${^1}S_0 \rightarrow {^3}P_1$ isotope shifts are shown in Table~\ref{table:isotopeshift}. 
The total systematic uncertainties are determined by adding the individual systematic uncertainties for each isotope in Table~\ref{table:intercombsystematics} in quadrature.
Our results are consistent with a previous measurement of the ${^{88}}$Sr-${^{86}}$Sr isotope shift, which reported a value of $163817.4 \pm 0.2$ kHz~\cite{ferrari03}.

\begin{table*}[htbp]
\begin{center}
\begin{tabular}{c c c c c c c}
  \hline \hline	
  & \multicolumn{2}{c}{88-87} & \multicolumn{2}{c}{88-86} & \multicolumn{2}{c}{88-84} \\
  \cline{2-7}
  Systematic Shift (kHz) & 88 & 87 & 88 & 86 & 88 & 84   \\ \hline
  Density & $0.8 \pm 1.6$ & $-3.8 \pm 1.2$ & $0.2 \pm 0.3$ & $-0.9 \pm 0.8$ & $0.4 \pm 0.9$ & $-2.3 \pm 0.9$ \\
  Recoil & $4.7 \pm (<0.1)$ & $4.7 \pm (<0.1)$ & $4.7 \pm (<0.1)$ & $4.8  \pm (<0.1)$ & $4.7  \pm (<0.1)$ & $4.9  \pm (<0.1)$ \\
  AC Stark & $51 \pm 5$ & $42 \pm 22$ & $53 \pm 5$ & $51 \pm 5$ & $50 \pm 5$ & $51 \pm 5$   \\
  Thermal & $-22 \pm 4$ & $-17 \pm 4$ & $-22 \pm 4$ & $-8 \pm 4$ & $-22 \pm 4$ & $-21 \pm 4$ \\
  2nd Order Zeeman & $-2.8 \pm (<0.1)$ & $0.0 \pm (<0.1)$ & $-2.8 \pm (<0.1)$ & $-2.8 \pm (<0.1)$ & $-2.8 \pm (<0.1)$ & $-9.1 \pm (<0.1)$ \\
  Probe Power & $3.5 \pm 1.6$ & $1.3 \pm 0.3$ & $3.5 \pm 1.6$ & $3.5 \pm 1.6$ & $3.6 \pm 1.6$ & $3.6 \pm 1.6$ \\
  Probe Duration & $3.4 \pm 3.3$ & $3.3 \pm 1.3$ & $3.4 \pm 3.3$ & $3.4 \pm 3.3$ & $3.4 \pm 3.3$ & $3.4 \pm 3.3$ \\
  \hline
  Total & $39 \pm 8$ & $31 \pm 22$ & $40 \pm 7$ & $51 \pm 7$ & $37 \pm 7$ & $31 \pm 7$ \\
  \hline \hline
\end{tabular}
\caption{Systematic frequency shifts and one standard deviation uncertainties for the ${^1}S_0 \rightarrow {^3}P_0$ transition. 
The three columns for ${^{88}}$Sr correspond to three independent isotope shift measurements.
Uncertainties indicate one standard deviation.
\label{table:clocksystematics}}
\end{center}
\end{table*}

\section{Measurement of the ${^1}S_0 \rightarrow {^3}P_0$ isotope shifts}
\begin{figure}[htbp]
\includegraphics[width=0.49\columnwidth]{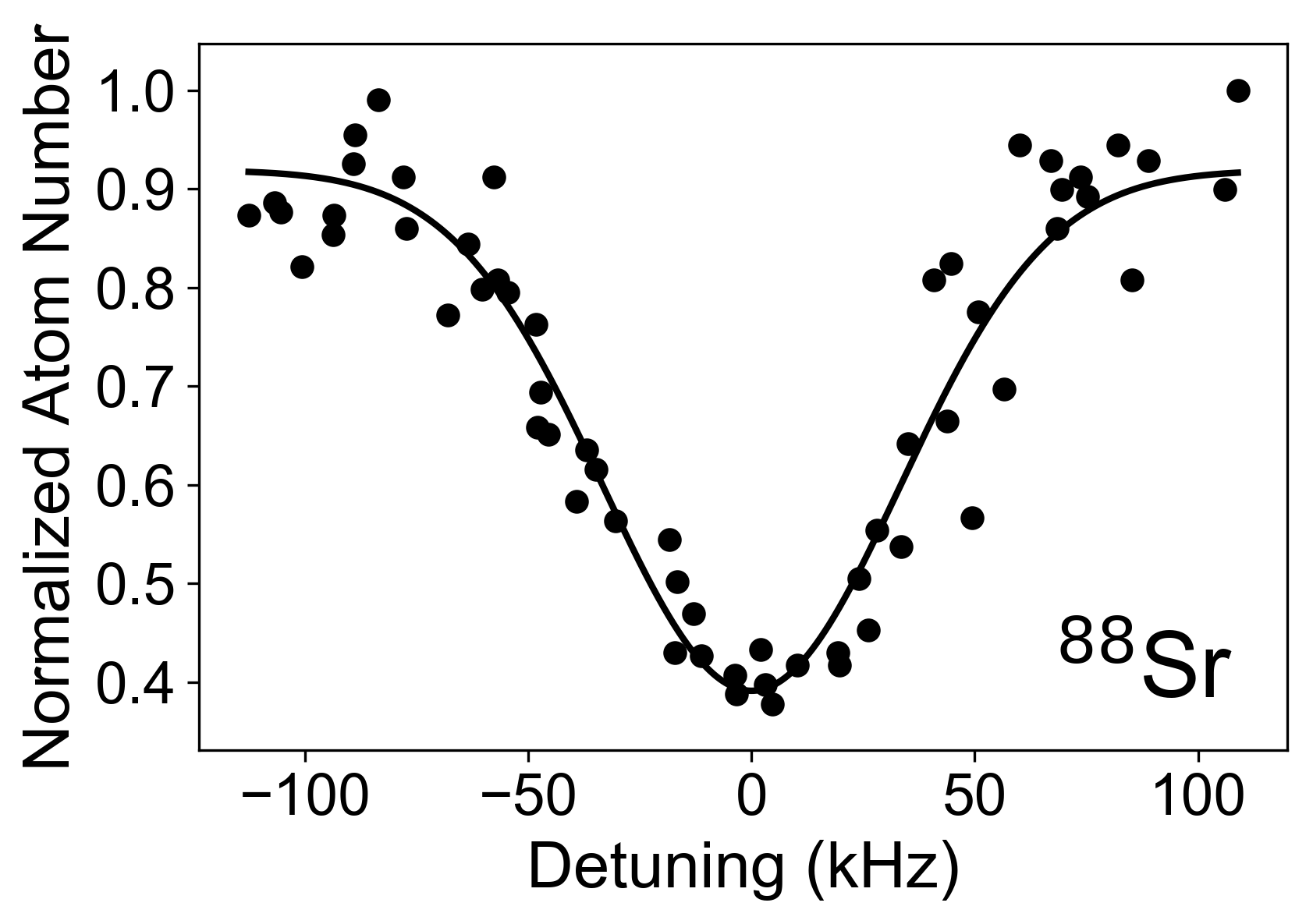}
\includegraphics[width=0.49\columnwidth]{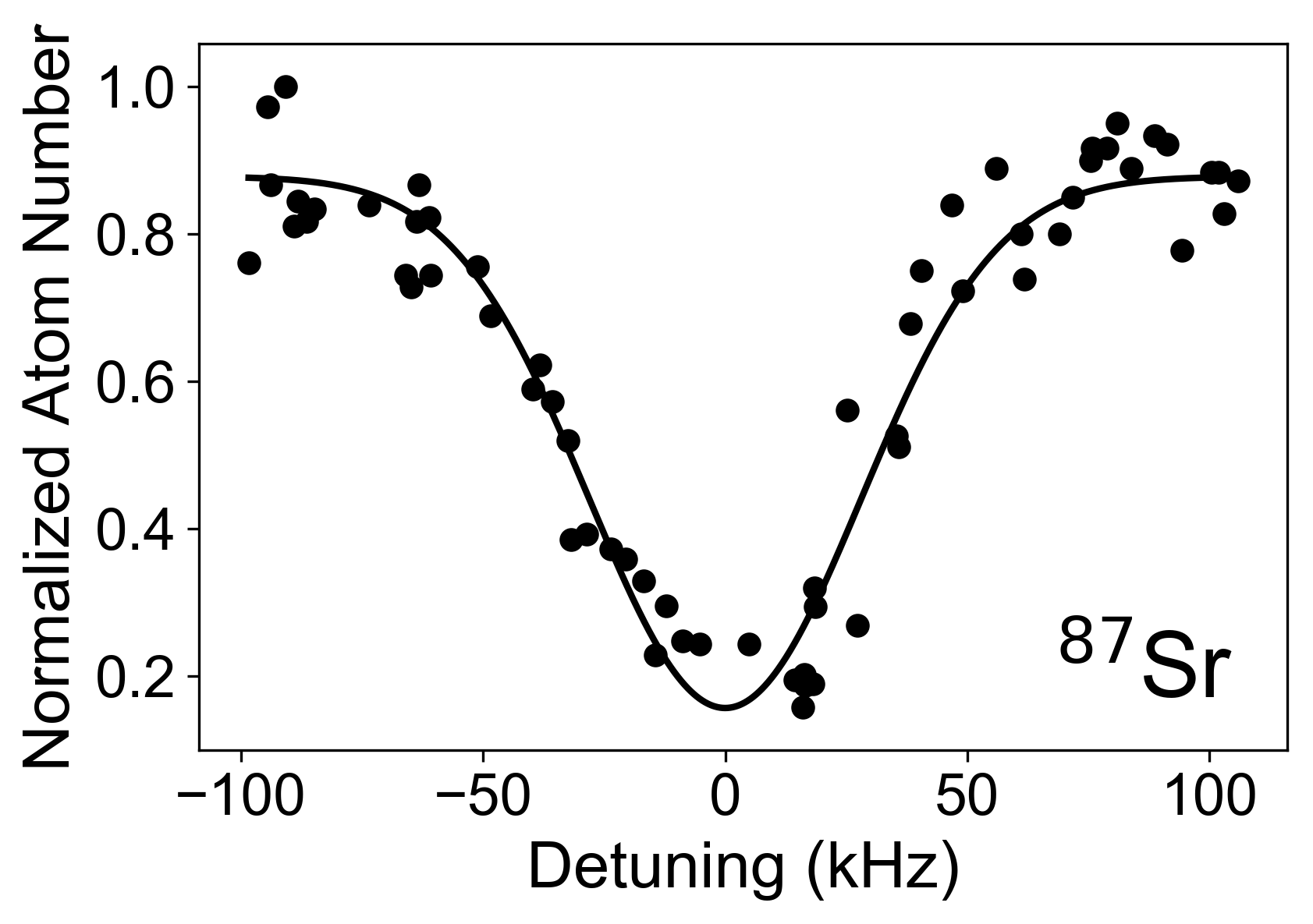}
\includegraphics[width=0.49\columnwidth]{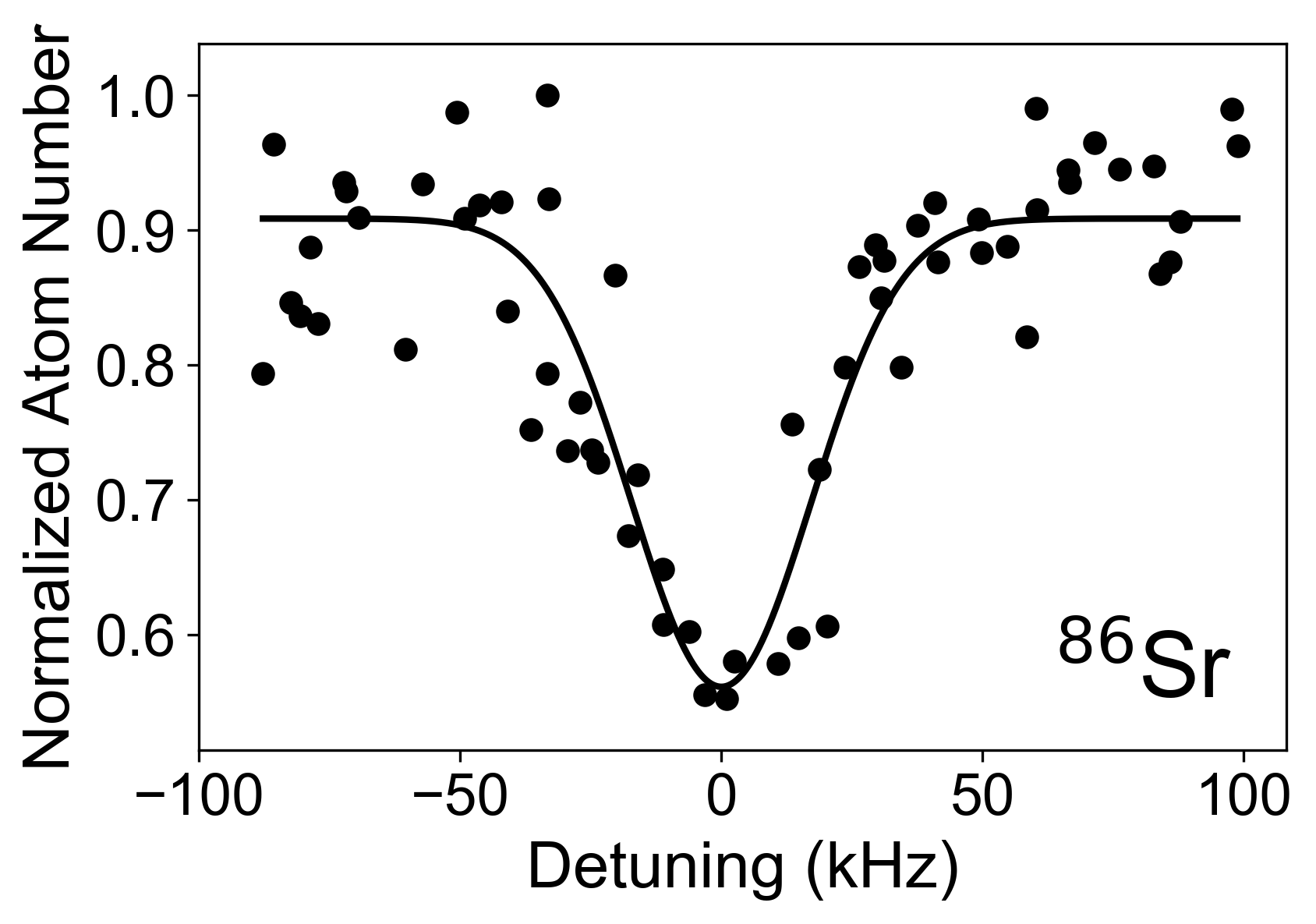}
\includegraphics[width=0.49\columnwidth]{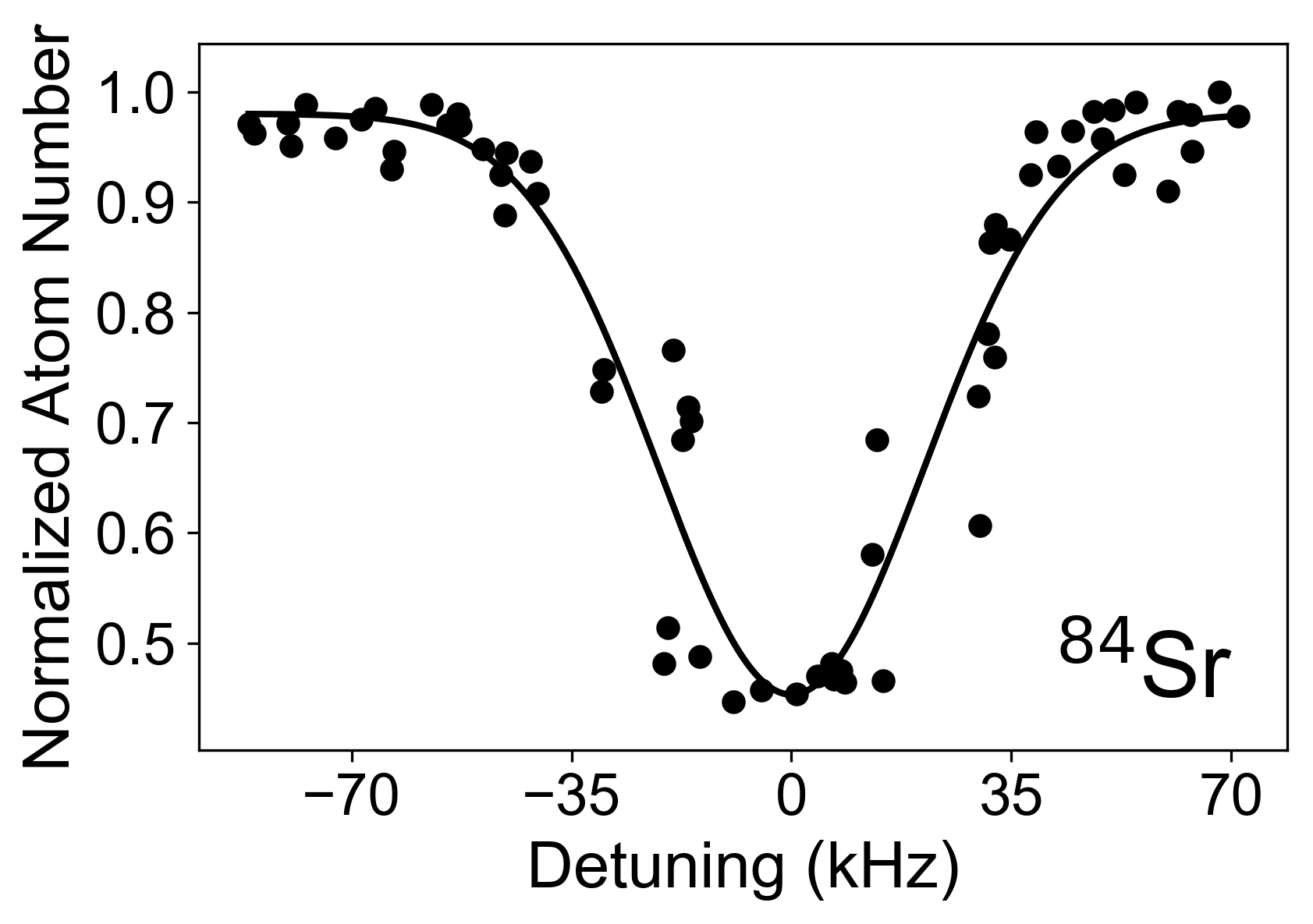}
\caption{\label{fig:scans} Spectroscopy of the ${^1}S_0 \rightarrow {^3}P_0$ transition for each strontium isotope. The normalized atom number is shown as a function of the laser detuning. The solid line is a Gaussian fit to the data.}
\end{figure}

The procedure for measuring the 698-nm transition differed from the measurement of the  ${^{1}}S_0 \rightarrow {^{3}}P_1$ intercombination-line transition in several key ways. Since the clock transition is strictly forbidden by angular momentum considerations for the bosonic isotopes, a much larger field was necessary to induce a transition in these isotopes. For $^{88}$Sr and  $^{86}$Sr a magnetic field of
$10.96 \pm 0.02$~mT ($109.6 \pm 0.2$~G)  was used, and
$19.79 \pm 0.05$~mT ($197.9 \pm 0.5$~G) was used for $^{84}$Sr.
For measurements of $^{87}$Sr, which is weakly allowed due to hyperfine mixing, we applied zero magnetic field.
For all isotopes, the 698-nm spectroscopy pulse was applied for 2~s with typical peak intensities of $0.87$ W/cm$^2$ for the even isotopes and 0.12 W/cm$^2$ for the odd isotope ($I_{\rm sat} \approx 0.4$ pW/cm$^2$).
These values were chosen to ensure approximately 50\% atom loss.
Atom loss was induced by the light scattering and subsequent recoil of the 698-nm light, which ejected the atoms out of the trap.
Representative line shapes for the ${^1}S_0 \rightarrow {^3}P_0$ transitions are show in Fig.~\ref{fig:scans} for each isotope.

The spectroscopy beam was aligned in the horizontal plane at an angle of approximately $45^\circ$ with respect to the ODT, and was focused onto the atoms with a $1/e^2$ waist of $330$ $\mu$m in the horizontal direction and $460$ $\mu$m in the vertical direction.
The beam was linearly polarized parallel to the magnetic field. 
Finally, because of the long interrogation time needed for sufficient atom loss (and therefore sufficient signal to noise), we were unable to apply the stroboscopic procedure used to measure the ${^1}S_0 \rightarrow {^3}P_1$ transitions, resulting in large AC Stark shifts from the trapping beam.
Due to the modified experimental procedure for the clock transition, additional systematic shifts included: thermal shifts, second order Zeeman shifts, and spectroscopy pulse shifts. 

For the ${^1}S_0 \rightarrow {^3}P_0$ transition, the dominant systematic effects were the AC Stark shift and what we call the thermal shift. 
The AC Stark shift arises from the differential polarizability of the ${^1}S_0$ and ${^3}P_0$ states at 1064~nm. 
The thermal shift arises from the inhomogeneous broadening and shift from the thermal motion in the intensity distribution of the ODT.
Experimentally, the AC Stark shift was determined by measuring the resonance frequency as a function of the ODT intensity. 
However, varying the intensity of the ODT also varied the trap depth, which in turn varied the temperature of the atomic cloud. 
This led to additional shifts in the resonance frequency due to both the Doppler shift and the inhomogeneous differential AC Stark shift.
To distinguish the effects of the AC Stark shifts from the thermal shifts we took the thermal average using the Maxwell-Boltzmann distribution and modeled the scattering process from the spectroscopy pulse~\cite{borkowski14,ciurylo04} (see supplemental material~\cite{supp}). As shown in Table~\ref{table:clocksystematics}, the experimentally determined values for the AC Stark shift for each isotope agree with each other within the uncertainty. Therefore the AC stark shift is common mode and cancels to a high degree. 

For even isotopes, the systematic shift for the first order Zeeman effect is zero since we probe a $J=0 \rightarrow J'=0$ transition with no hyperfine structure. To determine the second order Zeeman shifts for the ${^1}S_0 \rightarrow {^3}P_0$ transitions we used our calibrated magnetic field measurements and the known second order Zeeman shifts~\cite{taichenachev06}, which are identical for all even isotopes. 
Spectroscopy of ${^{87}}$Sr was performed at zero magnetic field, and so the Zeeman shift was well below other systematic effects~\cite{boyd07}.

The last systematic effects evaluated for the clock transition were related to the spectroscopy laser, occurring due to the relatively long probe time (2~s) and high peak intensities ($0.87$ W/cm$^2$). 
To measure these systematics, the transition frequency was measured as a function of both pulse power and duration, and the shift was extrapolated to zero. Finally, the density shift and recoil shift were obtained using the same procedure as described for the ${^1}S_0 \rightarrow {^3}P_1$ transition.

The final values for the isotope shift of the clock transition, including systematic corrections, are shown in Table~\ref{table:isotopeshift}. The systematic shifts are summarized in Table~\ref{table:clocksystematics}. The total systematic uncertainties for the clock transitions in Table~\ref{table:isotopeshift} are determined by adding the individual systematic uncertainties for each isotope in Table~\ref{table:clocksystematics} in quadrature. Comparing to prior measurements of the ${^{88}}$Sr-${^{87}}$Sr isotope shift which were all approximately 
$62188 \pm (< 1)$ kHz~\cite{baillard07,akatsuka08,takano17}, 
our result of 
$62171 \pm 24$ kHz
is consistent to well within one standard deviation.

\section{King plot analysis}
We performed a King plot analysis using our measured values of the isotope shifts, including the first measurements of the ${^{88}}$Sr-${^{86}}$Sr and ${^{88}}$Sr-${^{84}}$Sr isotope shifts for the clock transition.
A King plot analysis is a systematic approach to quantitatively and visually analyze isotope shifts of different atomic transitions referenced to the same isotope by relating the isotope shifts between different transitions~\cite{king84}.
This is a function of the mass and field shift constants, which are independent of the isotopes and depend only on the transitions under consideration~\cite{king63}.
Specifically, the isotope shifts between isotopes of mass numbers $A$ and $A'$ on two transitions $i$ and $j$
can be written
\begin{align}
\mu_{A,A'} \delta \nu_i^{A,A'} = K_i - \frac{F_i}{F_j} K_j + \frac{F_i}{F_j} \mu_{A,A'} \delta \nu_j^{A,A'},
\label{KingLinearity}
 \end{align}
 where $1/\mu_{A,A'} = 1/m_{A'} - 1/m_A$ is the inverse mass constant, 
$m_{A}$ is the mass of isotope $A$~\cite{emsley95}, 
$K_{i}$ is a constant associated with the mass shift of transition $i$, 
$F_{i}$ is the field shift constant for transition $i$, and 
$\delta \nu_{i}^{A,A'} = \nu_{i}^{A'} - \nu_{i}^{A}$ is the isotope shift between isotopes $A$ and $A'$ on transition $i$~\cite{king84,shi17}. 
For our particular analysis, we have $A = 88$, and  $A' \in \{87, 86, 84\}$, $i \equiv {^1}S_0 \rightarrow {^3}P_0$ at 698 nm, and $j \equiv {^1}S_0 \rightarrow {^3}P_1$ at 689 nm.
An important point to note is that Eq.~\ref{KingLinearity} describes a linear relationship between isotope shifts of different transitions. 

\begin{figure}[htbp]
\includegraphics[width=1\columnwidth]{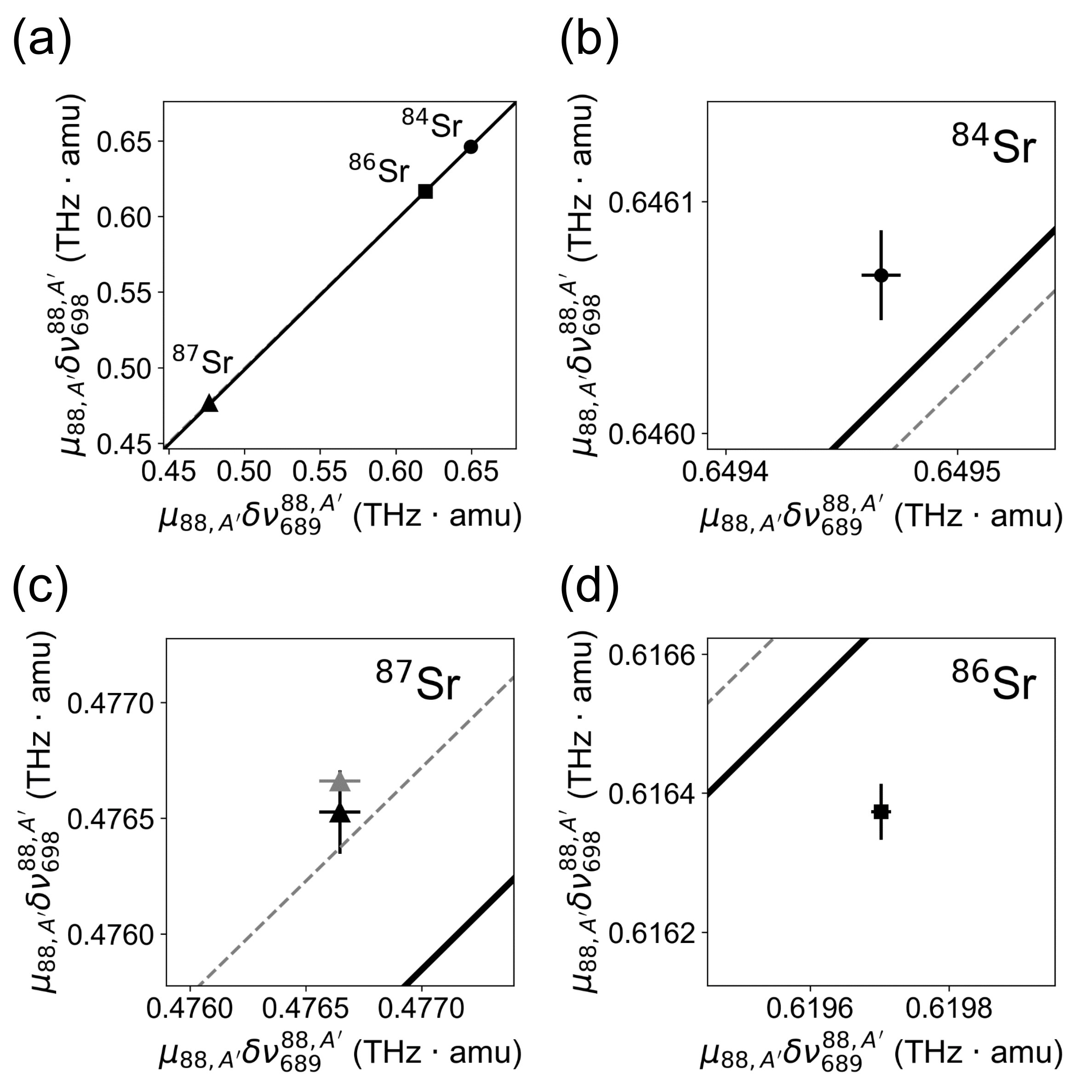}
\caption{\label{fig:king_plot} King plot of the measured strontium isotope shifts. 
(a) Linear fit to the three points derived from the six isotope shift measurements. 
Solid black line is a fit using all six of our measured isotope shifts, and the dashed gray line is a fit by replacing our measured $^{88}$Sr-$^{87}$Sr 698-nm transition isotope shift with the value from Ref.~\cite{takano17}, which is more precise than our measurement. 
The black points are derived from our measurements, and the gray point is using the $^{88}$Sr-$^{87}$Sr 698-nm transition isotope shift from Ref.~\cite{takano17}.
The fits are weighted by the uncertainties of each point. 
Error bars and the difference between the $^{87}$Sr points derived from our measurement and from Ref.~\cite{takano17} are not visible at this scale. 
(b)-(d) Close up of each point in (a) with error bars shown.}
\end{figure}

The King plot for our measured isotope shifts is shown in Fig.~\ref{fig:king_plot}. 
A linear fit to all three points weighted by their uncertainties leads to a field shift constant ratio of 
$F_{698}/F_{689} = 0.987 \pm 0.008$
and 
$K_{698} - \frac{F_{698}}{F_{689}} K_{689} = 5.20 \pm 5.31$ GHz$\cdot$amu, 
where the statistical and systematic uncertainties are added in quadrature. 
We have also performed a linear fit by replacing our measurement of the $^{88}$Sr-$^{87}$Sr 698-nm transition isotope shift with the more precise value from Ref.~\cite{takano17}. This leads to values of 
$F_{698}/F_{689} = 0.981 \pm 0.005$
and 
$K_{698} - \frac{F_{698}}{F_{689}} K_{689} = 8.56 \pm 3.45$ GHz$\cdot$amu,
which are consistent with values obtained using our measurement of the $^{88}$Sr-$^{87}$Sr 698-nm transition isotope shift.

Since there is some uncertainty in deriving the frequency for ${^{87}}$Sr due to the hyperfine structure, we also fit the data after excluding this point to obtain a field shift constant ratio of 
$F_{698}/F_{689} = 0.998 \pm 0.002$ 
and 
$K_{698} - \frac{F_{698}}{F_{689}} K_{689} = -1.87 \pm 1.03$ GHz$\cdot$amu 
where the uncertainties are propagated from the uncertainties of each point for both axes. 
Compared to this two-point linear fit, the $^{88}$Sr-$^{87}$Sr 689-nm isotope shift we determined would have to increase by 136.2 kHz to become consistent with a linear King plot. 
Given that our data points with their uncertainties lie well outside of the straight line fit to all three points, the results in Fig.~\ref{fig:king_plot} suggest a possible nonlinear contribution to Eq.~\ref{KingLinearity}, or may indicate significant uncertainties in the determination of the center-of-mass of the $^{87}$Sr $^{3}P_1$ hyperfine structure. 
In particular, our data indicates a nonlinearity using the nonlinearity measure defined in Ref.~\cite{berengut18}. Future theoretical and experimental studies should help to explain our observations, including better calculations of the hyperfine mixing within the $^{3}P$ states and a prediction of the King plot slope.

\section{Conclusions}
In summary, we have presented the first spectroscopy of the ${^1}S_0 \rightarrow {^3}P_0$ clock transition in $^{86}$Sr and $^{84}$Sr, and reported their isotope shifts relative to $^{88}$Sr.
In conjunction with improved measurements of the intercombination line isotope shifts, we performed a King plot analysis and extracted constants related to the field and mass shifts.
Hyperfine effects in $^{87}$Sr complicate this analysis, but the experimental precision permitted by these two narrow optical transitions make it a rich testbed to benchmark state-of-the-art theory. Furthermore, it has been suggested that a comparison of isotope shifts between neutral and ionic strontium could set stringent limits on new physics~\cite{frugiuele17,berengut18}. 
However, an improved theory, accounting for our observed nonlinearity would be essential. 
Alternatively, one could also perform this measurement with the radioactive bosonic isotope $^{90}$Sr (half-life of approximately 29 years~\cite{schrader04}) to avoid complications due to the hyperfine structure.

Future improvements on the measured frequencies will be possible by applying techniques successfully used with state-of-the-art strontium optical clocks, such as the use of magic-wavelength dipole traps to minimize the differential AC stark shift~\cite{katori03,takamoto05} and optical lattices to suppress motional broadening and recoil shifts~\cite{ido03}.
These advances should further suppress statistical and systematic errors on both transitions, allowing measurements with fractional uncertainties down to the level of $10^{-18}$~\cite{bloom14,nicholson15,campbell17}.
Our results, combined with other recent measurements of isotope shifts in Ca$^+$~\cite{knollmann19} and Sr$^+$~\cite{manovitz19},  will further help to refine refine atomic structure calculations and constrain new physics.

\begin{acknowledgments}
We thank Luis Orozco, Marianna Safronova, and Charles Clark for fruitful discussions and Nicholas Mennona for experimental assistance. This work was partially
supported by the U.S. Office of Naval Research and the NSF through the Physics Frontier Center at the Joint Quantum Institute.
\end{acknowledgments}

\appendix*

\section{Modeling the inhomogeneous broadening of the clock transition}
In general, the AC Stark shift is different for different atomic states due to state-dependent polarizabilities.  The exception to this is if one operates the dipole trap at specific laser wavelengths typically referred to as the ``magic wavelength'' where the ground and excited states experience the same AC Stark shifts. For strontium atoms, the magic wavelength is 813 nm for the 698-nm clock transition and 914 nm for the 689-nm intercombination transition~\cite{boyd07,bloom14}. In our experiment, the optical dipole trap uses 1064-nm laser light, a wavelength where the two states, ${^1}S_0$ and ${^3}P_0$, have different polarizabilities. This leads to inhomogeneous broadening which must be accounted for.  The resulting lineshape is further complicated by the temperature of our atomic samples. Here we describe our method for modeling and accounting for this inhomogeneous broadening due to both the differential AC Stark shift and the thermal shift.

We model the inhomogeneous broadening process using a semi-classical treatment of atom loss from the trap due to the spectroscopy pulse~\cite{ciurylo04}.
We can model the atom loss from the spectrocopy pulse after some probe time, by calculating the loss rate coefficient $K$.
The time-dependent atom number in the presence of the spectroscopy pulse is governed by the differential equation
\begin{align}
\frac{d N}{d t} &= -K(\delta\omega,I,T,U_\text{trap}) N, \label{Eq:rateeq}
\end{align}
where the loss rate coefficient $K$ is a function of the laser detuning $\delta\omega = \omega_\text{laser} - \omega_0$ ($\omega_{\rm laser}$ is the frequency of the probe laser and $\omega_0$ is the bare atomic resonance frequency), the probe laser intensity $I$, the atomic cloud temperature $T$, and the dipole trap potential $U_\text{trap}$. The loss rate is modeled to be proportional to an ensemble average of the scattering rate over all atoms in the trap. 

The scattering rate can be written~\cite{metcalf99}
\begin{align}
\Gamma_\text{scat} &= \frac{\Gamma}{2}\left(\frac{s_0}{1+s_0+\left(2\Delta/\Gamma\right)^2}\right),
\end{align}
where $\Gamma$ is the transition linewidth, $\Delta$ is the effective detuning from resonance, $s_0 \equiv I/I_{\rm sat}$ is the on-resonance saturation parameter, $I$ is the excitation laser intensity, and $I_{\rm sat}$ is the saturation intensity.
We rearrange this expression, pulling out constant terms to write
\begin{align}
\Gamma_\text{scat} &\propto \frac{1}{\left(\Gamma'/2\right)^2 + \Delta^2},\label{eq:lineshape:scatt}
\end{align}
where $\Gamma' = \Gamma\sqrt{1+s_0}$ is the saturation-broadened linewidth.
For a thermal atom in a far-detuned optical dipole trap with a given phase space coordinate $(\mathbf r, \mathbf p)$, $\Delta$ can be written
\begin{align}
\Delta &= \delta\omega - \frac{\mathbf p\cdot \mathbf k}{m} - \left( U_e(\mathbf r) - U_g(\mathbf r) \right),\label{eq:detuning:scatt}
\end{align}
where the term $\delta\omega - \mathbf p\cdot \mathbf k/m$ is the Doppler-shifted laser frequency, $\mathbf p$ is the atomic momentum vector, $\mathbf k$ is the probe laser wavevector, $m$ is the atomic mass, and $U_e(\mathbf r) - U_g(\mathbf r)$ is the differential AC Stark shift which arises from different polarizabilities between the states $e$ and $g$. 
Note that in the treatment here we neglect all other systematic frequency offsets which do not depend on position, since these appear simply as frequency offsets and do not cause any inhomegeneous effects. 
We also neglect gravity in our model since the atoms are tightly confined in this direction.
We can approximate the trapping potential for the far-detuned optical trap as a parabola and write
\begin{align}
U_e(\mathbf r) - U_g(\mathbf r) &= U_{e,0} + \frac{1}{2}m \bar\omega_e^2 \mathbf r^2 - U_{g,0} - \frac{1}{2}m \bar\omega_g^2 \mathbf r^2 \\
&= U_{e,0} - U_{g,0} + \frac{1}{2}m\left(\bar\omega_e^2 - \bar\omega_g^2\right)\mathbf r^2 \\
&= U_{e,0} - U_{g,0} + \frac{1}{2}m\bar\omega_g^2 \mathbf r^2\left(\frac{\bar\omega_e^2}{\bar\omega_g^2} - 1\right) \\
&= \Delta U_0 + \Delta U_\text{trap}(\mathbf r),
\end{align}
where $\bar{\omega}_{g} (\bar{\omega}_e)$ is the geometric mean of the ground (excited) state trap frequencies in all three dimensions and the trap is effectively spherical in these coordinates.
Since $\bar{\omega}_i \propto \sqrt{\alpha_i}$, where $\alpha_i$ is the AC polarizability of state $i\in \{g,e\}$, we find
\begin{align}
\Delta U_\text{trap}(\mathbf r) &= U_\text{trap}(\mathbf r)\left(\frac{\alpha_e}{\alpha_g} - 1\right),
\end{align}
where $U_\text{trap} =m\bar\omega_g^2 \mathbf r^2/2$.
For 1064-nm light, with $g$ the ${^1}S_0$ state and $e$ the ${^3}P_0$ state, we compute $\alpha_e/\alpha_g \approx 0.7$. 
This can also be written as a re-scaling of the trap potential, such that
\begin{align}
\Delta U_\text{trap}(\mathbf r) &= \alpha U_\text{trap}(\mathbf r),
\end{align}
with $\alpha = (\alpha_e/\alpha_g - 1) \approx -0.295$.
Note that operating the dipole trap at the magic wavelength would lead to $\alpha_e = \alpha_g$, which means $\alpha = 0$, and therefore the spatial dependence would drop out of Eq.~\ref{eq:detuning:scatt}.

We now turn our attention to solving for the loss rate coefficient $K$ by taking an ensemble average over the scattering rate expressed in Eq.~\ref{eq:lineshape:scatt} using the detuning defined in Eq.~\ref{eq:detuning:scatt}.
Because we are interested in deriving a lineshape function which can be fit to experimentally measured atom loss data, we ignore normalization and overall constant terms which can be condensed into a single fit parameter.
Taking the ensemble average of Eq.~\ref{eq:lineshape:scatt} leads to
\begin{widetext}
\begin{align}
K \propto \int d^3\mathbf r~e^{-U_{\text{trap}}(\mathbf r)/(k_B T)}~\int d^3\mathbf p~e^{-{\mathbf p}^2/(2mk_B T)}~\left[\frac{1}{\left(\Gamma'/2\right)^2 + \left(\delta\omega - \Delta U_0 - \mathbf p \cdot \mathbf k/m - \alpha U_\text{trap}(\mathbf r)\right)^2}\right]\,,\label{Eq:ensembleAve}
\end{align}
\end{widetext}
where we have taken an integral over phase space $(\mathbf r, \mathbf p)$ weighted by the Boltzmann factor.
Here, $k_B$ is the Boltzmann constant.

We wish to make this dimensionless to easily work in a numerical fitting routine with experimental data.
Focusing on the integral {$d^3\mathbf p = dp_x dp_y dp_z$} first, we can choose $\hat p_z$ to point along $\mathbf k$.
Thus, $\mathbf p \cdot \mathbf k = p_z k$, and the Boltzmann factor can be rewritten
\begin{align}
e^{-{\mathbf p}^2/(2mk_B T)} &= e^{-(p_x^2 +p_y^2)/(2mk_B T)} e^{-p_z^2/(2mk_B T)}
\end{align} 
The integral over $p_x$ and $p_y$ now factors out, and can be brought into an overall scale factor.
We define the dimensionless variable $y \equiv p_z/\sqrt{2mk_B T}$. After defining $\beta \equiv k\sqrt{2 k_B T/m}$, this becomes $p_z k/m = \beta y$.
In convenient units, for ${^{88}}$Sr and $2\pi/k=698$ nm, we get $\beta/2\pi =$ 19.7 kHz$\cdot \sqrt{T}$ with $T$ measured in $\mu$K.
This parameterization of $y$ serves to scale the momentum $p_z$ to the most probable momentum at a given temperature.

Putting it all together, the integral from Eq.~\ref{Eq:ensembleAve} becomes
\begin{widetext}
\begin{align}
K \propto \int d^3\mathbf r~e^{-U_{\text{trap}}(\mathbf r)/(k_B T)}~\int dy~e^{-y^2}~\left[\frac{1}{\left(\Gamma'/2\right)^2 + \left(\delta\omega - \Delta U_0 - \beta y - \alpha U_\text{trap}(\mathbf r)\right)^2}\right].
\end{align}
\end{widetext}
With regards to the integral over $\mathbf r$, since we have scaled the trap to be effectively spherical, we can write $U_\text{trap}(\mathbf r) = f(r^2)$.
Thus, we can pull the angular integral from $d^3\mathbf r \equiv r^2\sin\theta dr~d\theta~d\phi$ into an overall constant, leaving just the integral in $r$ given by
\begin{widetext}
\begin{align}
K \propto \int r^2 dr~e^{-r^2 m \bar\omega_g^2/(2k_B T)}~\int dy~e^{-y^2}~\left[\frac{1}{\left(\Gamma'/2\right)^2 + \left(\delta\omega - \Delta U_0 - \beta y - \alpha m \bar\omega_g^2 r^2/2 \right)^2}\right]\,,
\end{align}
\end{widetext}
where we replaced $U_\text{trap}(\mathbf r)$ with its explicit form $m\bar\omega_g^2r^2/2$.

Defining the dimensionless variable $x\equiv r\sqrt{m\bar\omega_g^2/(2k_B T)} = r~(\bar\omega_g k/\beta)$, which scales $r$ by the ratio of the trap potential energy to the thermal energy $k_B T$, we can rewrite the integral as
\begin{widetext}
\begin{align}
K \propto \int dx~x^2 e^{-x^2}~\int dy~e^{-y^2}~\left[\frac{1}{\left(\Gamma'/2\right)^2  + \left(\delta\omega - \Delta U_0 - \beta y - \left(\frac{\alpha m }{2 k^2}\right)\beta^2 x^2 \right)^2}\right]\,,\label{Eq:finalint}
\end{align}
\end{widetext}
where for our system, $\alpha m/(2 k^2) \approx \SI{-2.52e-6}{\second}$.

Returning to Eq.~\ref{Eq:rateeq}, we use our expression Eq.~\ref{Eq:finalint} for $K$ to solve for atom number and obtain
\begin{align}
\frac{N(\tau)}{N(0)} = e^{-K \tau}\,,
\label{eq:lineshape:numfit}
\end{align}
which can be used as an integral function to fit the four parameters $\{a, \left(\omega_0 + \Delta U_0(I_{\rm trap}) \right), \Gamma',\beta\}$, where $a$ is an overall normalization factor for $K$, in a least-squares minimization routine. We keep the $\Delta U_0(I_{\rm trap})$ term explicit and highlight its dependence on the optical dipole trap laser intensity $I_{\rm trap}$. We use this expression to extract the AC Stark shift systematic correction.

\begin{figure}
    \includegraphics[width=0.99\columnwidth]{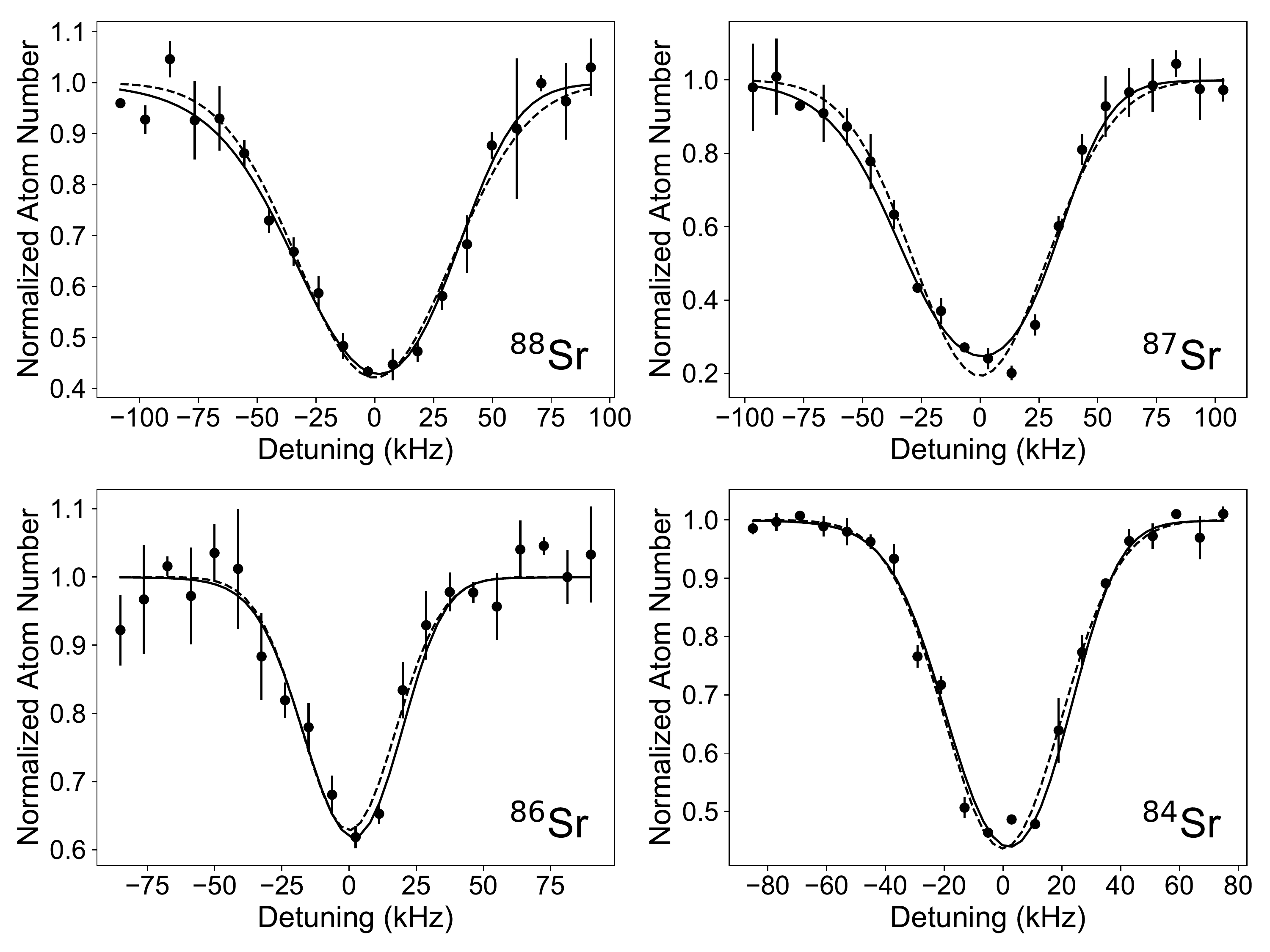}
\caption[Detailed lineshape curves for \SI{698}{\nano\meter} clock transition]{
Lineshape curves for the 698-nm clock transition. The curves include a Gaussian model (dashed line) and a full lineshape model (solid line) fit to the averaged data points (circle points).
In both cases, the fit error on the centroid is roughly \SI{1}{\kilo\hertz}, however the full lineshape model fits a different $\omega_0$ which varies as a function of temperature and is red of the Gaussian line center by up to 20 kHz.
This is attributable to the thermal distribution of atoms in a dipole trap with inhomogeneous AC Stark shifts.
\label{fig:isotope:lineshapeScan}
}
\end{figure}

\begin{figure}
%\begin{figure}[H]
    \includegraphics[width=0.99\columnwidth]{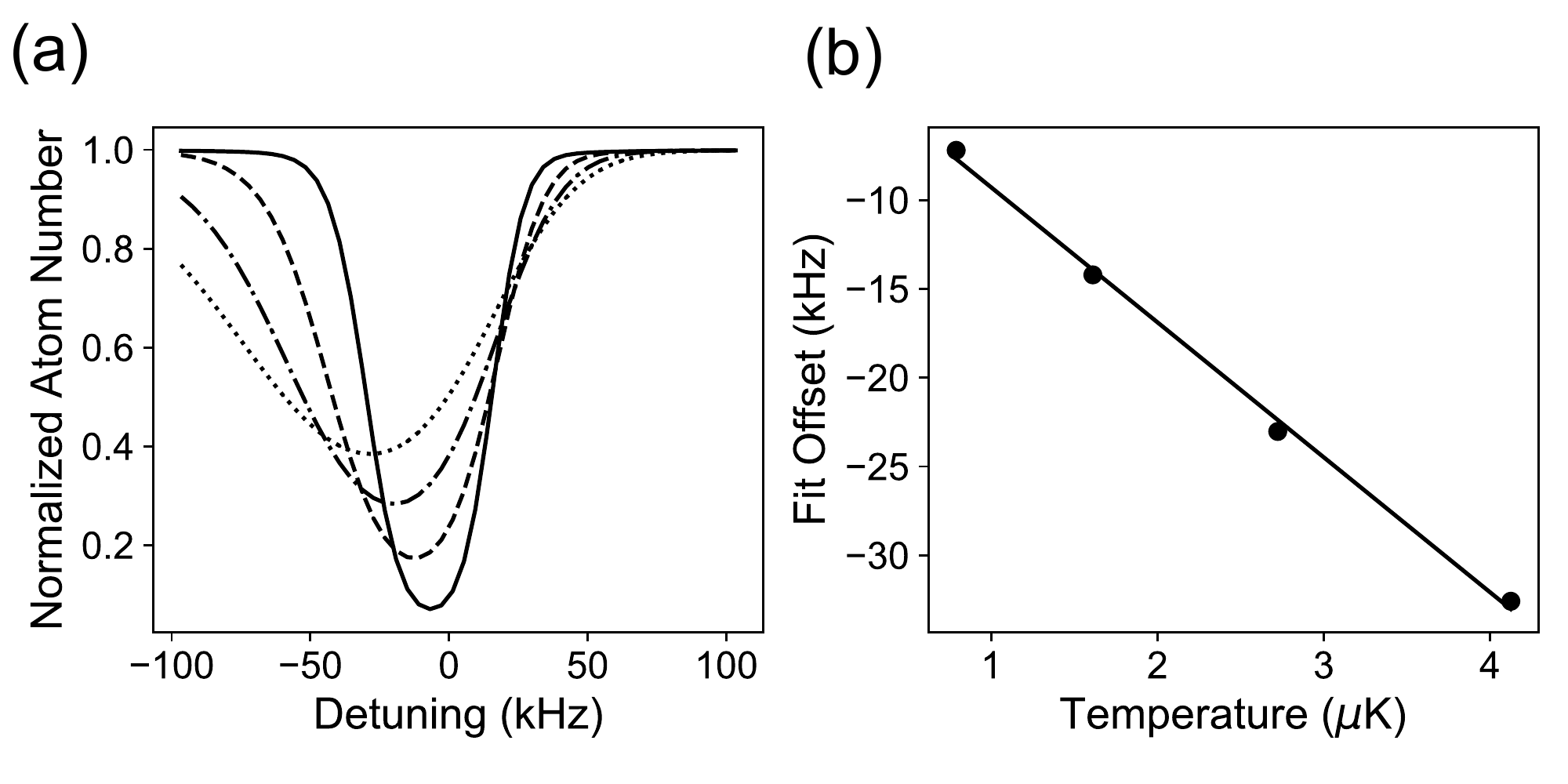}
    \caption[Thermal line pulling for clock transition]{
Effects of thermal line shift on the clock transition. 
(a) Lineshape simulations as a function of temperature with 0.79 $\mu$K (solid line), 1.6 $\mu$K (dashed line), 2.7 $\mu$K (dash-dotted line), and 4.1 $\mu$K (dotted line). (b) Systematic offset to the Gaussian fitted center as a function of temperature and a linear fit to the data extracted from the simulation in (a).
\label{fig:lineshape:pulling}
}
\end{figure}

Note that in theory, the integral in Eq.\ref{Eq:finalint} ranges over the entire real line. In our numerical implementation, we truncate these integrals at finite values. In our experiment, we typically have $U_{\rm trap} \sim 160$ kHz, and so $U_{\rm trap}/(k_B T) \sim 8$ and we take the position integral out to five times the thermal energy scale. Since the integrand is convolved by a gaussian, continuing the integration further in the wings contributes only marginally to the final value, and the truncation does not change the result above other uncertainties. The ratio $U_{\rm trap}/(k_B T) \sim 8$ also allows us to approximate the trap as harmonic.

As an example, we perform a fit using Eq.~\ref{eq:lineshape:numfit} to the loss spectra shown in Fig.~\ref{fig:isotope:lineshapeScan}.
It is difficult to visually differentiate the quality of the fit between the full integral lineshape and a simple Gaussian model, but there is a non-negligible thermal line shift from a full accounting of the lineshape as is evident in the fit parameters.
To account for this systematic shift, we numerically simulate the systematic Gaussian fit offset as a function of temperature, and find it to be $-7.6 \pm 0.3$ kHz/$\mu$K, as shown in Fig.~\ref{fig:lineshape:pulling}. With this result and measured temperatures of \{$2.9$ $\mu$K, $2.2$ $\mu$K, $1.1$ $\mu$K $2.7$ $\mu$K\}, we obtain a systematic frequency shifts of \{$-22 \pm 4$ kHz, $-17 \pm 4$ kHz, $-8 \pm 4$ kHz, $-21 \pm 4$ kHz\} for \{${^{88}}$Sr, ${^{87}}$Sr, ${^{86}}$Sr, ${^{84}}$Sr\} respectively.


\begin{thebibliography}{99}

\bibitem{king84}
W. H. King, \textit{Isotope Shifts in Atomic Spectra}, Physics of Atoms and Molecules (Plenum, New York, 1984).

\bibitem{flambaum18}
V. V. Flambaum, A. J. Geddes, and A. V. Viatkina, Phys. Rev. A \textbf{97}, 032510 (2018).

\bibitem{griffith81}
 J. A. R. Griffith \textit{et al}., J. Phys. B: At. Mol. Phys. \textbf{14}, 2769 (1981).
 
 \bibitem{dammalapati09}
 U. Dammalapati, S. De, K. Jungmann, and L. Willmann, Eur. Phys. J. D \textbf{53}, 1 (2009).

\bibitem{shi17}
C. Shi \textit{et al}., Appl. Phys. B \textbf{123}, 2 (2017).

\bibitem{naze15}
C. Naz\'{e}, J. G. Li, and M. Godefroid, Phys. Rev. A \textbf{91}, 032511 (2015).

\bibitem{frugiuele17}
C. Frugiuele, E. Fuchs, G. Perez, and M. Schlaffer, Phys. Rev. D \textbf{96}, 015011 (2017).

\bibitem{berengut18}
J. C. Berengut \textit{et al}., Phys. Rev. Lett. \textbf{120}, 091801 (2018).

\bibitem{stellmer14}
S. Stellmer, F. Schreck, and T. C. Killian, Annual Review of Cold Atoms and Molecules, Volume 2, Chapter 1, World Scientific (2014).

\bibitem{daley11}
A. J. Daley, Quantum Inf. Process., \textbf{10}, 865 (2011).

\bibitem{he19}
C. He, E. Hajiyev, Z. Ren, B. Song, and G.-B. Jo, J. Phys. B: At. Mol. Opt. Phys. \textbf{52}, 102001 (2019).

\bibitem{katori99}
H. Katori, T. Ido, Y. Isoya, and M. Kuwata-Gonokami, Phys. Rev. Lett. \textbf{82}, 1116 (1999).

\bibitem{bloom14}
B. J. Bloom \textit{et al}., Nature \textbf{506}, 71 (2014).

\bibitem{nicholson15}
T. L. Nicholson \textit{et al}., Nat. Commun. \textbf{6}, 6896 (2015).

\bibitem{campbell17}
S. L. Campbell \textit{et al}., Science \textbf{358}, 90 (2017).

\bibitem{taichenachev06}
A. V. Taichenachev, V. I. Yudin, C. W. Oates, C. W. Hoyt, Z. W. Barber, and L. Hollberg, Phys. Rev. Lett. {\bf 96}, 083001 (2006).

\bibitem{baillard07}
X. Baillard \textit{et al}., Opt. Lett. \textbf{32}, 1812 (2007).

\bibitem{akatsuka08}
T. Akatsuka, M. Takamoto, and H. Katori, Nat. Phys. \textbf{4}, 954 (2008).

\bibitem{morzynski15}
P. Morzy\'{n}ski \textit{et al}., Sci. Rep. \textbf{5}, 17495 (2015).

\bibitem{takano17}
T. Takano, R. Mizushima, and H. Katori, Appl. Phys. Express \textbf{10}, 072801 (2017).

\bibitem{ludlow15}
A. D. Ludlow, M. M. Boyd, J. Ye, E. Peik, and P. O. Schmidt, Rev. Mod. Phys. \textbf{87}, 637 (2015).

\bibitem{cook12}
E. C. Cook, P. J. Martin, T. L. Brown–Heft, J. C. Garman, and D. A. Steck, Rev. Sci. Instrum. {\bf 83}, 043101 (2012).

\bibitem{appel09}
J. Appel, A. MacRae, and A. I. Lvovsky, Meas. Sci. Technol. \textbf{20}, 055302 (2009).

\bibitem{drever83}
R. W. P. Drever \textit{et al}., Appl. Phys. B \textbf{31}, 97, (1983).

\bibitem{black01}
E. D. Black, Am. J. Phys. \textbf{69}, 79 (2001).

\bibitem{thorpe08}
J. I. Thorpe, K. Numata, and J. Livas, Opt. Express {\bf 16}, 15980 (2008).

\bibitem{nist_disclaimer}
The identification of commercial products in this paper is for information
only and does not imply recommendation or endorsement by the National
Institute of Standards and Technology.

\bibitem{barker15}
D. S. Barker, B. J. Reschovsky, N. C. Pisenti, and G. K. Campbell, Phys. Rev. A \textbf{92}, 043418 (2015).

\bibitem{mukaiyama03}
T. Mukaiyama, H. Katori, T. Ido, Y. Li, and M. Kuwata-Gonokami, Phys. Rev. Lett. 90, 113002 (2003).

\bibitem{borkowski14}
M. Borkowski, P. Morzy\'{n}ski, R. Ciury\l{}o, P. S. Julienne, M. Yan, B. J. DeSalvo, and T. C. Killian, Phys. Rev. A \textbf{90}, 032713 (2014).

\bibitem{reschovsky18}
B. J. Reschovsky \textit{et al}., arXiv:1808.06507 [physics.atom-ph].

\bibitem{atomicStructure}
G. K. Woodgate, \textit{Elementary Atomic Structure} (Clarendon Press, Oxford, 2000).

\bibitem{hfcoeff}
We determined $|A| = 260085 \pm 2$ kHz and $|B|~=~35667~\pm~21$ kHz, consistent with previous results~\cite{putlitz63}.

\bibitem{beloy08}
K. Beloy, A. Derevianko, and W. R. Johnson, Phys. Rev. A \textbf{77}, 012512 (2008).

\bibitem{ido03}
T. Ido and H. Katori, Phs. Rev. Lett. \textbf{91}, 053001 (2003).

\bibitem{ferrari03}
G. Ferrari, P. Cancio, R. Drullinger, G. Giusfredi, N. Poli, M. Prevedelli, C. Toninelli, and G. M. Tino, Phys. Rev. Lett. \textbf{91}, 243002 (2003).

\bibitem{ciurylo04}
R. Ciury\l{}o, E. Tiesinga, S. Kotochigova, and and P. S. Julienne, Phys. Rev. A \textbf{70}, 062710 (2004).

\bibitem{supp}
See Supplementary Material.

\bibitem{boyd07}
M. M. Boyd, T. Zelevinsky, A. D. Ludlow, S. Blatt, T. Zanon-Willette, S. M. Foreman, and J. Ye, Phys. Rev. A \textbf{76}, 022510 (2007).

\bibitem{king63}
W. H. King, J. Opt. Soc. Am. \textbf{53}, 638 (1963).

\bibitem{emsley95}
J. Emsley, \textit{The Elements}, Oxford Chemistry Guides (Oxford Univ. Press, New York, NY, 1995).

\bibitem{schrader04}
H. Schrader, Appl. Radiat. Isot. \textbf{60}, 317 (2004).

\bibitem{katori03}
H. Katori, M. Takamoto, V. G. Pal'chikov, and V. D. Ovsiannikov, Phys. Rev. Lett. \textbf{91}, 173005 (2003).

\bibitem{takamoto05}
M. Takamoto, F.-L. Hong, R. Higashi, and H. Katori, Nature \textbf{435}, 321 (2005).

\bibitem{knollmann19}
F. W. Knollmann, A. N. Patel, and S. C. Doret, arXiv:1906.04105 [physics.atom-ph].

\bibitem{manovitz19}
T. Manovitz, R. Shaniv, Y. Shapira, R. Ozeri, and N. Akerman, arXiv:1906.05770 [physics.atom-ph].

\bibitem{putlitz63}
G. zu Putlitz, Z. Phys. \textbf{175}, 543 (1963).



\bibitem{boyd07}
M. M. Boyd, PhD thesis, University of Coloardo, 2007.

\bibitem{metcalf99}
H. J. Metcalf and P. van der Straten, Laser Cooling and Trapping, (Springer, New York, 1999).

\end{thebibliography}
\end{document}